\documentclass[10pt,twocolumn,letterpaper]{article}

\usepackage{iccv}
\usepackage{booktabs}

\usepackage{multirow}

\definecolor{iccvblue}{rgb}{0.21,0.49,0.74}
\definecolor{cvprblue}{rgb}{0.21,0.49,0.74}
\usepackage[pagebackref,breaklinks,colorlinks,allcolors=cvprblue]{hyperref}

\def\confName{ICCV}
\def\confYear{2025}

\def\OURS{{POMATO}\xspace}
\def\bF{{\bf F}}
\def\bC{{\bf C}}
\def\bI{{\bf I}}

\def\bD{{\bf D}}

\def\bK{{\bf K}}

\def\bX{{\bf X}}

\def\bG{{\bf G}}

\def\bhead2{{$Head_2$}}
\def\bhead1{{$Head_1$}}

\def\matR{{\mathbb{R}}}

\usepackage{stfloats}
\usepackage{multirow,marvosym}
\usepackage{array}

\title{\OURS: Marrying Pointmap Matching with Temporal Motions \\ for Dynamic 3D Reconstruction}

\author{
Songyan Zhang$^{1*}$ \quad
Yongtao Ge$^{2,3*}$ \quad
Jinyuan Tian$^{2*}$ \quad
Guangkai Xu$ ^2$ \quad \\
Hao Chen$^2$\textsuperscript{\Letter} \quad
Chen Lv$ ^1 $ \quad
Chunhua Shen$ ^2 $ \quad
\\[0.24cm]
$ ^1 $ Nanyang Technological University, Singapore
~~~
$ ^2 $ Zhejiang University, China \\
~~~
$ ^3 $ The University of Adelaide, Australia
}

\begin{document}

\newcommand{\ours}{POMA{ }}
\newcommand{\duster}{DUSt3R}
\newcommand{\monster}{MonST3R}
\newcommand{\aftertab}{\vspace{-0.5em}}
\newcommand{\afterfig}{\vspace{-0.3em}}
\newcommand{\aroundeqn}{\vspace{0em}}
\newcommand{\aftertabcaption}{\vspace{0.5em}}
\newcommand{\beforetfigcaption}{\vspace{-1.2em}}

\twocolumn[{%
\renewcommand\twocolumn[1][]{#1}%

\maketitle

\includegraphics[width=0.99\linewidth]{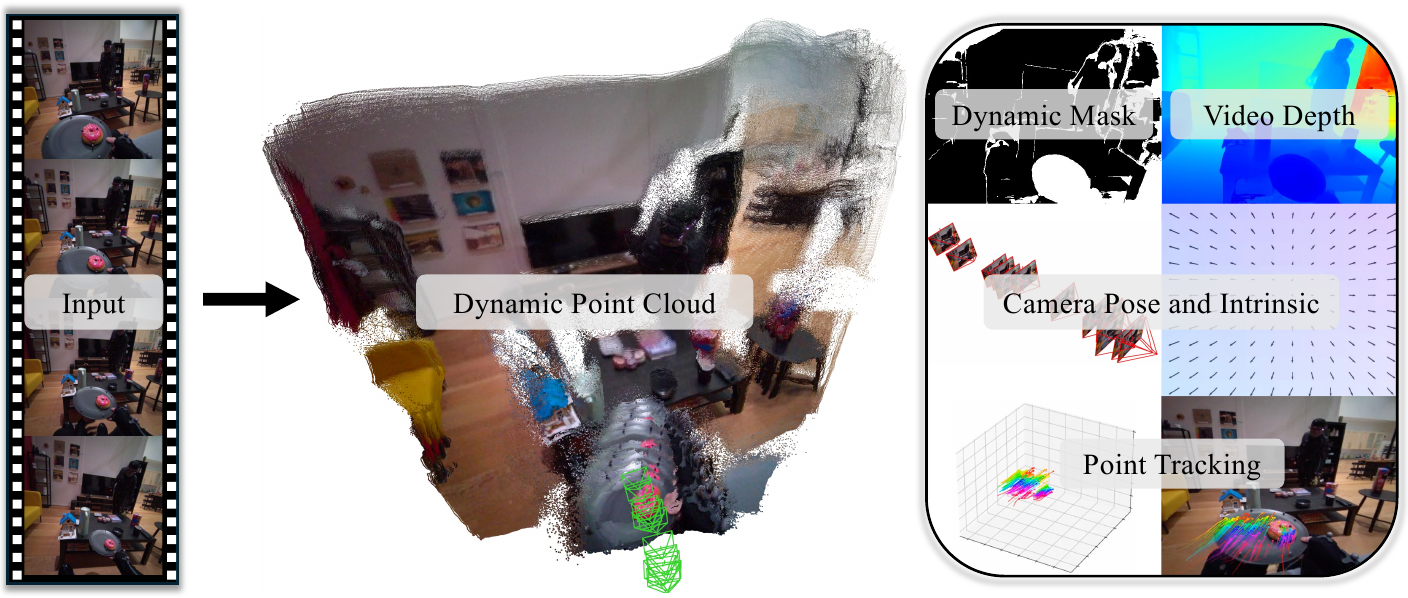}

\captionof{figure}{\textbf{3D reconstruction from an arbitrary dynamic video with \OURS.} Without relying on external modules, \OURS can directly perform 3D reconstruction along with temporal 3D point tracking and dynamic mask estimation.}
\label{fig:teaser}
\vspace{1em}
}]

\begin{NoHyper}
\let\thefootnote\relax\footnotetext{* Equal contribution. \textsuperscript{\Letter} Corresponding author. 
}
\end{NoHyper}

\begin{abstract}
Recent approaches to 3D reconstruction in dynamic scenes primarily rely on the integration of separate geometry estimation and matching modules, where the latter plays a critical role in distinguishing dynamic regions and mitigating the interference caused by moving objects.
Furthermore, the matching module explicitly models object motion, enabling the tracking of specific targets and advancing motion understanding in complex scenarios. 
Recently, the proposed representation of pointmap in DUSt3R suggests a potential solution to unify both geometry estimation and matching in 3D space, effectively reducing computational overhead by eliminating the need for redundant auxiliary modules.
However, it still struggles with ambiguous correspondences in dynamic regions, which limits reconstruction performance in such scenarios. In this work, we present \OURS, a unified framework for dynamic 3D reconstruction by marrying \textbf{PO}intmap \textbf{MA}tching with \textbf{\text{T}}emporal m\textbf{\text{O}}tion. Specifically, our method first learns an explicit matching relationship by mapping RGB pixels across different views to 3D pointmaps within a unified coordinate system. Furthermore, we introduce a temporal motion module for dynamic motions that ensures scale consistency across different frames and enhances performance in 3D reconstruction tasks requiring both precise geometry and reliable matching, most notably 3D point tracking. We show the effectiveness of our proposed POMATO by demonstrating the remarkable performance across multiple downstream tasks, including video depth estimation, 3D point tracking, and pose estimation. Code and models are publicly available at \url{https://github.com/wyddmw/POMATO}.

\end{abstract}    
\section{Introduction}
\label{sec:intro}
\begin{figure*}[!t]
    \centering
    \includegraphics[width=0.99\linewidth]{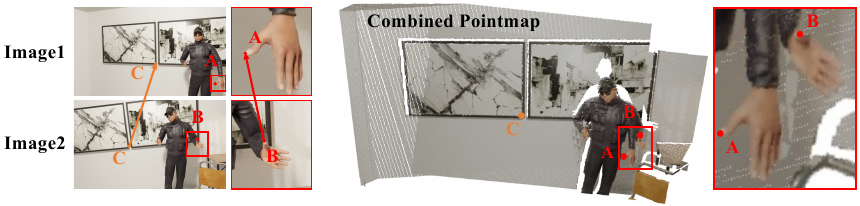}
    \caption{\textbf{Ambiguity in 3D point matching in dynamic scenes with DUSt3R.} Given representative corresponding pixels of background (orange) and moving foreground (red) in two different views, DUSt3R outputs a pair of 3D points within the same coordinate system. In static regions, identical pixels share the same 3D coordinates which provide an accurate matching relationship in 3D space, but in moving regions, the 3D coordinates are inconsistent for corresponding pixels across views, leading to ambiguous 3D matching relationships.}
    \vspace{-4mm}
    \label{fig:ambiguity}
\end{figure*}

Image-based 3D reconstruction is a fundamental task in computer vision with a wide range of applications including SLAM \cite{droidslam}, robotics~\cite{irshad2024neural, xu2024dgslam}, autonomous driving~\cite{zhao2024drive}, and novel view synthesis~\cite{hexplane}. While substantial progress has been achieved in static 3D reconstruction~\cite{metric3dv2, depthanythingv2, marigold, dust3r, mast3r}, dynamic scenes remain a major hurdle due to complexities like non-rigid motion and deformation, which may hamper the learning of local structure and camera motion, thereby complicating accurate 3D reconstruction for dynamic scenes. These scenarios require explicit modeling of both scene geometry and object motion. Moreover, downstream reconstruction tasks, such as 3D point tracking, demand precise geometry estimation and robust matching across views. To effectively distinguish dynamic regions, it is essential to establish reliable correspondences between different frames.
Some pioneering works have attempted to address dynamic motion by incorporating additional auxiliary matching modules, such as optical flow \cite{shapeofmotion, monst3r} or 2D tracking \cite{spatialtracker}. However, these approaches may suffer from domain gaps and accumulated errors between modules, limiting their effectiveness. A unified framework that seamlessly integrates geometry estimation and matching for dynamic 3D reconstruction remains a critical and underexplored challenge.

Recently, DUSt3R~\cite{dust3r} proposes a promising solution to address this challenge. It introduces the concept of a pointmap that assigns each pixel in an image to a corresponding 3D coordinate. The network utilizes a standard transformer-based encoder-decoder architecture and receives a pair of images as input. The system incorporates two parallel decoders to predict pointmaps for each view within the same coordinate system. However, this representation is limited to static matching and struggles in dynamic scenes, as illustrated in Fig.~\ref{fig:ambiguity}. 

To address this problem, we present \OURS, a unified network for dynamic 3D reconstruction 
by marrying \textbf{PO}intmap \textbf{MA}tching with \textbf{\text{T}}emporal m\textbf{\text{O}}tion.
We argue that with iterative cross-attention modules across different views, matching features are well preserved in the decoder tokens. We thus introduce an auxiliary pointmap matching head to learn explicit correspondences. Specifically, for each pixel in the second view, the pointmap matching head predicts the corresponding 3D coordinates of its counterpart in the first view, under the shared coordinate system. Our proposed pointmap-based matching representation enables the establishment of explicit correspondences in 3D space, which can be directly leveraged for motion analysis, especially the estimation of dynamic regions. Moreover, we further extend our \OURS to handle 4D video sequences by introducing a temporal motion module that enhances the learning of temporal motions. This motion module promotes scale consistency across different frames and improves performance in tasks where both accurate geometry and reliable matching are paramount, most notably 3D point tracking. Compared with recent temporal 3D reconstruction methods \cite{cut3r, spann3r} based on an autoregression manner where the previous frames are blocked from the recently added frames, our temporal motion module is based on the self-attention mechanism along the temporal dimension, facilitating a comprehensive interaction across all frames. Our \OURS is trained in a two-stage manner. In the first stage, we used pairwise input images to learn fundamental geometry and matching capabilities. In the second stage, we extend the input to sequential video frames and incorporate the temporal motion module, enabling the model to effectively capture motions over time.

Our contributions can be summarized in threefold: First, we propose a novel approach that unifies the fundamental geometry estimation and motion understanding for dynamic 3D reconstruction into a single network by incorporating the representation of pointmap matching. Second, we introduce a temporal motion module to facilitate the interactions of motion features along the temporal dimension, which significantly improves the performance in tasks where both accurate geometry and precise matching are required for video sequential input, most notably 3D point tracking. Third, we demonstrate promising performance on 3D vision tasks, including video depth estimation, 3D point tracking, and camera pose estimation.

\section{Related Work}
\label{sec:rel_work}

\noindent\textbf{Geometry estimation} refers to the process of determining the spatial properties and structures from different forms of visual data. Direct recovery of 3D geometry from a single RGB image is by nature an ill-posed problem. Many recent works~\cite{marigold, depthanythingv2, depthpro, metric3dv2} have tried to leverage strong pre-trained models to learn generalizable depthmaps from large-scale real and synthetic datasets to solve ambiguities. For example, Marigold~\cite{marigold}, Geowizard~\cite{geowizard}, and GenPercept~\cite{genpercept} aim at leveraging the generative priors from pre-trained diffusion models by finetuning them on synthetic datasets. Depthanything V2~\cite{depthanythingv2} proposes to estimate scale-and-shift invariant disparity map by finetuning DINOV2~\cite{dinov2} model on synthetic datasets and large-scale pseudo labels. Depth Pro~\cite{depthpro} further propose a FOV head to estimate the metric depthmap from a single image without relying on camera intrinsics as input. Due to the scale ambiguity in the monocular depth estimation models, ChronoDepth~\cite{chronodepth}, DepthCrafter~\cite{depthcrafter}, and Depth-any-video~\cite{depthanyvideo} proposes to learn temporal consistent depthmaps by leveraging the priors from a video generative model, \ie SVD~\cite{svd}.
In another line of the research, multi-view stereo reconstruction (MVS) methods seek to reconstruct visible surfaces from multiple viewpoints. Traditional MVS~\cite{furukawa2015multi} and SfM pipelines break the reconstruction pipeline into several sub-problems, \eg, feature extraction~\cite{detone2018superpoint}, image matching~\cite{barath2023affineglue,lightglue}, triangulation, and bundle adjustment~\cite{chen2019bundle}. The chain is complicated and accumulates noise for every single step, thus often resulting in unsatisfactory performance in complex real-world scenes. Recognizing the limitations of previous MVS methods, seminal work DUSt3R~\cite{dust3r} proposes 3D pointmaps representation, and trains a network from large-scale data to regress the dense and accurate pointmaps from a pair of images. The camera intrinsics and relative camera poses can be implicitly inferred from the two-view pointmaps. However, it still can not handle reconstruction for dynamic scenes. MonST3R \cite{monst3r} directly finetuned the original DUSt3R model upon synthetic datasets that contain dynamic scenes.

\noindent\textbf{Motion representation.} Optical flow is a commonly used representation for 2D motion. RAFT~\cite{raft} is a representative work for pairwise optical flow estimation, which employs a 4D cost volume and recurrently estimates the optical flow. Some follow-up methods further extend it to multi-frame (3-5 frames) settings, which is still insufficient for long-range tracking. To resolve the problem, Particle Video~\cite{particlevideo} represent video motion by using a set of particles. Each particle is an image point sample with a long-duration trajectory and other properties. Particle videos have two key advantages over optical flow: (1) persistence through occlusions, and (2) multi-frame temporal context. Some recent works, PIPs~\cite{pips}, TAPIR~\cite{tapir} and Cotracker~\cite{cotracker} have renewed interest in this representation and show promising long-term 2D point tracking results. Recognizing the advantage of point representation, SpatialTracker~\cite{spatialtracker} lifts the 2D points into 3D and performs tracking in the 3D space. Though it can handle occlusions and enhance 3D tracking accuracy, it still relies on a separate monocular depth estimator, which prevents it performing 3D point tracking in an end-to-end fashion.

\noindent\textbf{Multi-view dynamic reconstruction.} Our work is closely connected to multi-view dynamic 3D reconstruction techniques. Early works~\cite{russell2014video,ranftl2016dense} take the straightforward idea that first pre-segment the scene into different regions, each corresponding to a single rigid part of an object, then apply the rigid-SfM technique to each of the regions. Some of the recent Neural Radiance Fields (NeRF) ~\cite{nerf} and Gaussian Splatting ~\cite{3dgs} based methods have achieved state-of-the-art results. However, most of these methods require simultaneous multi-view video inputs or require predefined templates~\cite{humanrf}.
Shape of motion~\cite{shapeofmotion}, proposes a new dynamic scene representation to represent the dynamic scene as a set of persistent 3D Gaussians, and optimize the representation from a monocular video by leveraging monocular depth estimation priors and 2D track estimates across frames.
\section{Method}

\begin{figure*}[!t]
    \centering
    \includegraphics[width=0.99\linewidth]{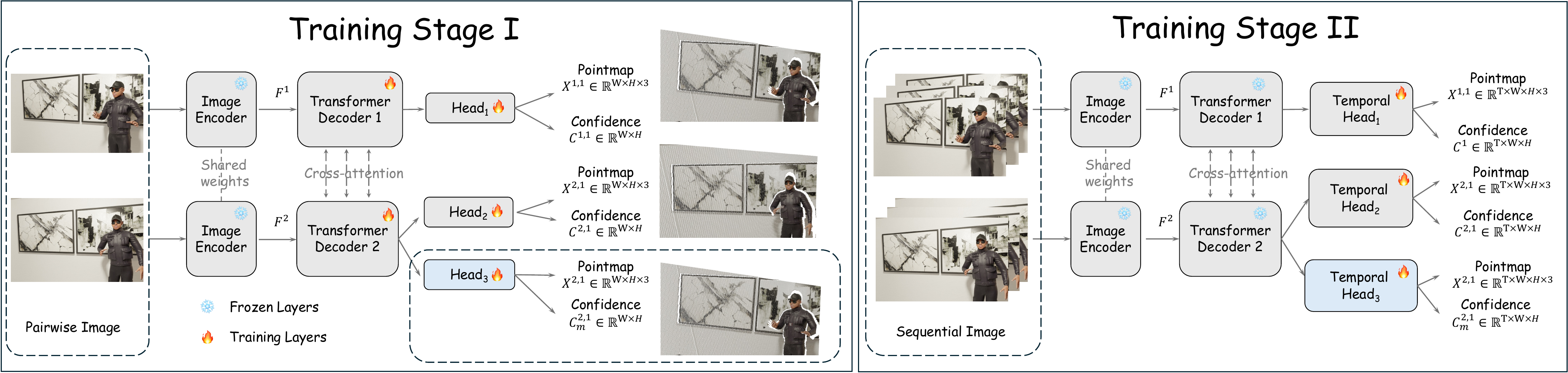}
    \caption{\textbf{Overview of our training pipeline.} (1) Stage I: build upon \duster{}~\cite{dust3r} architecture, we introduce a third regression point-matching head: Head$_3$, which is in parallel to Head$_2$ for explicit pointmap matching in 3D space. For each pixel in the second view, the output pointmap coordinate is the 3D point map of the corresponding pixel in the first view. (2) Stage II: we introduce a temporal fusion module in three heads that enables multi-style sequential input for learning temporal motions.}
    \vspace{-4mm}
    \label{fig:pipeline}
\end{figure*}

\subsection{Preliminary}

The overview of our \OURS is demonstrated in Fig.\ref{fig:pipeline}. We adopt the definition of pointmap $\bX \in \matR ^{H\times W \times 3}$ in DUSt3R~\cite{dust3r} as a dense 2D field of 3D points where each point corresponds to its respective RGB pixel. Given a pair of input images $\bI^1, \bI^2 \in \matR^{H \times W \times 3}$ from two different views, a weight-sharing ViT first extracts the corresponding features $\bF^1, \bF^2$ for each view. Two parallel branches are employed to decode the geometric structures and enhance the feature alignment via cross-attention in decoder modules, following a regression head to estimate pointmaps $\bX^{1,1}, \bX^{2,1} \in \matR^{H \times W \times 3}$ along with a confidence map $\bC^{1,1}, \bC^{2,1} \in \matR^{H \times W}$ for each image view. Generally, $\bX^{n,m}$ indicates the pointmap $\bX^n$ from camera $n$ expressed in camera $m$'s coordinate frame, which is obtained by a rigid transformation:

\begin{align}
    \bX^{n, m} = {\bf P}_m{\bf P}_n^{-1} h(\bX^n),
\end{align}
where ${\bf P}_m, {\bf P}_n \in \matR^{3\times 4}$ are world-to-camera poses for camera $m$ and camera $n$, respectively, and $h(\bX^{n})$ is a homogeneous mapping for the 3D coordinate in camera coordinate of camera $n$.

The task for Decoder 1 and its regression head estimate the 3D points for $\bI^1$ in its own coordinate system while Decoder 2 and its regression head are responsible for estimating pixel-wise 3D coordinates for $\bI^2$ in $\bI^1$'s coordinate system after a rigid transformation of global rotation and translation. In the following contents, we will first introduce our \OURS with pairwise input images and then extend it to the video sequence input with our temporal motion module.
\subsection{Pointmap Matching with Pairwise Input}

As discussed before, the definition of $\bX^{2,1}$ depicts a rigid camera transformation that is ambiguous to reflect explicit matching relationships for dynamic regions. To tackle this, we propose to formulate an explicit pointmap matching $\bX^{2,1}_m \in \matR ^{H\times W \times 3}$ that maps dense RGB pixels of $\bI^2$ to 3D coordinates of corresponding pixels in $\bI^1$ under the first image's coordinate system. Given a 2D query pixel at $(x_2, y_2)$ in $\bI^2$ and its corresponding pixel at $(x_1, y_1)$ in $\bI^1$, the matched pointmap at $(x_2, y_2)$ in $\bI^2$ is: 
\begin{align}
    \bX^{2,1}_m(x_2, y_2) = \bX^{1,1}(x_1, y_1),
    \label{eq:def_pointmap_matching}
\end{align}
where $(x, y)$ indicates the coordinates of 2D grid. For the representative dynamic point (red) in Fig. \ref{fig:ambiguity}, the pointmap matching result is the 3D coordinate of point A in the coordinate system of the first image. As shown in Fig. \ref{fig:pipeline}, $\bX^{2,1}_m$ and $\bX^{1,1}$ are supposed to match perfectly in 3D space on the premise of neglecting occluded regions. We argue that the set of decoder tokens from the second branch preserves abundant matching information with iterative cross-attentions, so we introduce a matching head with the same architecture of Head$_1$ and Head$_2$. The supervision for pointmap matching $\bX^{2,1}_m$ still follows the 3D regression loss which is defined as the Euclidean distance: 

\begin{equation}
\mathcal{L}_\text{m} = \left\Vert \frac{1}{z_m}{\bX^{2,1}_m}  - \frac{1}{\bar{z}_m}{\bar{\bX}^{2,1}}_m \right\Vert,
\vspace{-1mm}
\label{eq:regression}
\end{equation}
where $\bar{\bX}^{2,1}_m$ is the ground truth pointmap matching, which can be obtained following Eq. \ref{eq:def_pointmap_matching} on the 2D tracking dataset with the depth and camera information. $z_m, \bar{z}_m$ are the same norm factor defined in DUSt3R. The matching confidence $\bC^{2,1}_m$ is also learned following the confidence loss for Head$_1$ and Head$_2$ within valid regions:

\begin{equation}
\mathcal{L}_\text{mconf} = \bC^{2,1}_m \mathcal{L}_\text{m} - \alpha \text{log} \bC^{2,1}_m
\vspace{-1mm}
\label{eq:matching_confidence}
\end{equation}

The final loss $\mathcal{L}$ of our \OURS for pairwise input is a combination of predefined DUSt3R loss $\mathcal{L}_{\text{DUSt3R}}$, matching loss $\mathcal{L}_{\text{m}}$, and matching confidence loss $\mathcal{L}_{\text{mconf}}$. When training our \OURS for pairwise input images at the first stage, the parameters in the encoder are frozen.

\subsection{Dynamic Mask Estimation}\label{sec:dynamic_mask}
Taking advantage of the explicit pointmap matching head, our \OURS can directly perform dynamic mask estimation without introducing an assistant module such as the optical flow model, getting rid of the additional computation cost and the potential domain gap. For an image pair $\{\bI^i, \bI^j\}$ along with the estimation of $\bX^{j,i}$ from Head$_2$ and $\bX^{j,i}_m$ from Head$_3$, the dynamic mask $\bD^{j,i}$ can be obtained by comparing the difference between $\bX^{j,i}$ and $\bX^{j,i}_m$:
\begin{align}
    \bD^{j,i} = ||\bX^{j,i}_m - \bX^{j,i}|| > \alpha,
    \label{eq:dynamic_mask_eq}
\end{align}
where $\alpha$ is a dynamic threshold defined as $3 \times \text{median}(||\bX^{j,i}_m - \bX^{j,i}||)$. The explicit dynamic mask can be incorporated into the global alignment process to minimize the interference of moving objects for pose estimation and 3D reconstruction. Details on the incorporation of dynamic masks for global alignment are provided in the supplementary materials.

\begin{figure}[!t]
    \centering
    \includegraphics[width=0.8\linewidth]{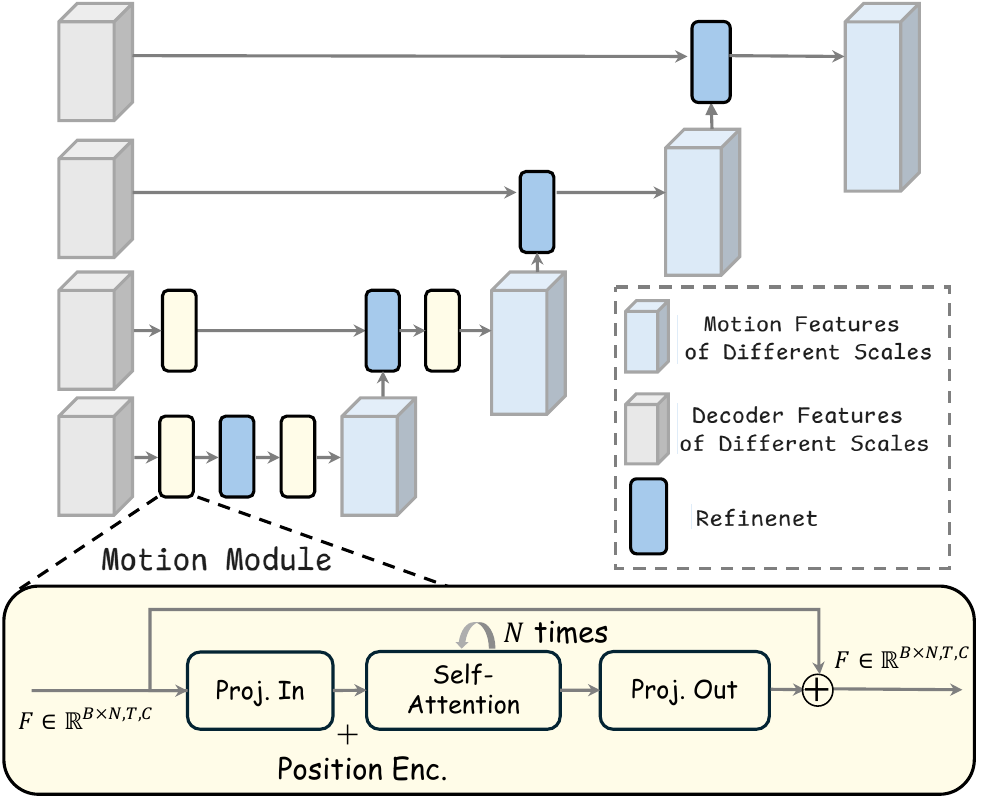}
    \caption{\textbf{Architecture of our temporal motion module.} We insert a transformer-based motion module (in shallow yellow) into the vanilla DPT~\cite{dpt} head to enhance the temporal consistency.}
    \vspace{-4mm}
    \label{fig:temporal_module}
\end{figure}

\subsection{Temporal Motion Module}\label{sec:motion_module}
With the fundamental capability of geometric estimation and pointmap matching for pairwise images, we follow \cite{videodepthanything-v2} and extend our \OURS to 4D video sequences by inserting a transformer-based motion module into the vanilla DPT head to construct the "temporal DPT head", which is illustrated in Fig.\ref{fig:temporal_module}. For a set of decoder tokens $\bG \in \matR^{B, T, N, C}$ where $B, T, N, C$ represent the batch size, window length of a video sequence, token number, and token dimension, respectively, we merge the token number dimension into the batch axis and apply the motion module which consists of two blocks of standard multi-head self-attention modules and feed-forward networks along the temporal dimension $T$. To reduce the computation cost, the temporal motion modules are applied to features of low resolution.

\subsection{Downstream Temporal Tasks}\label{sec:temporal_tasks}
\begin{figure}[!t]
    \centering
    \includegraphics[width=1.0\linewidth]{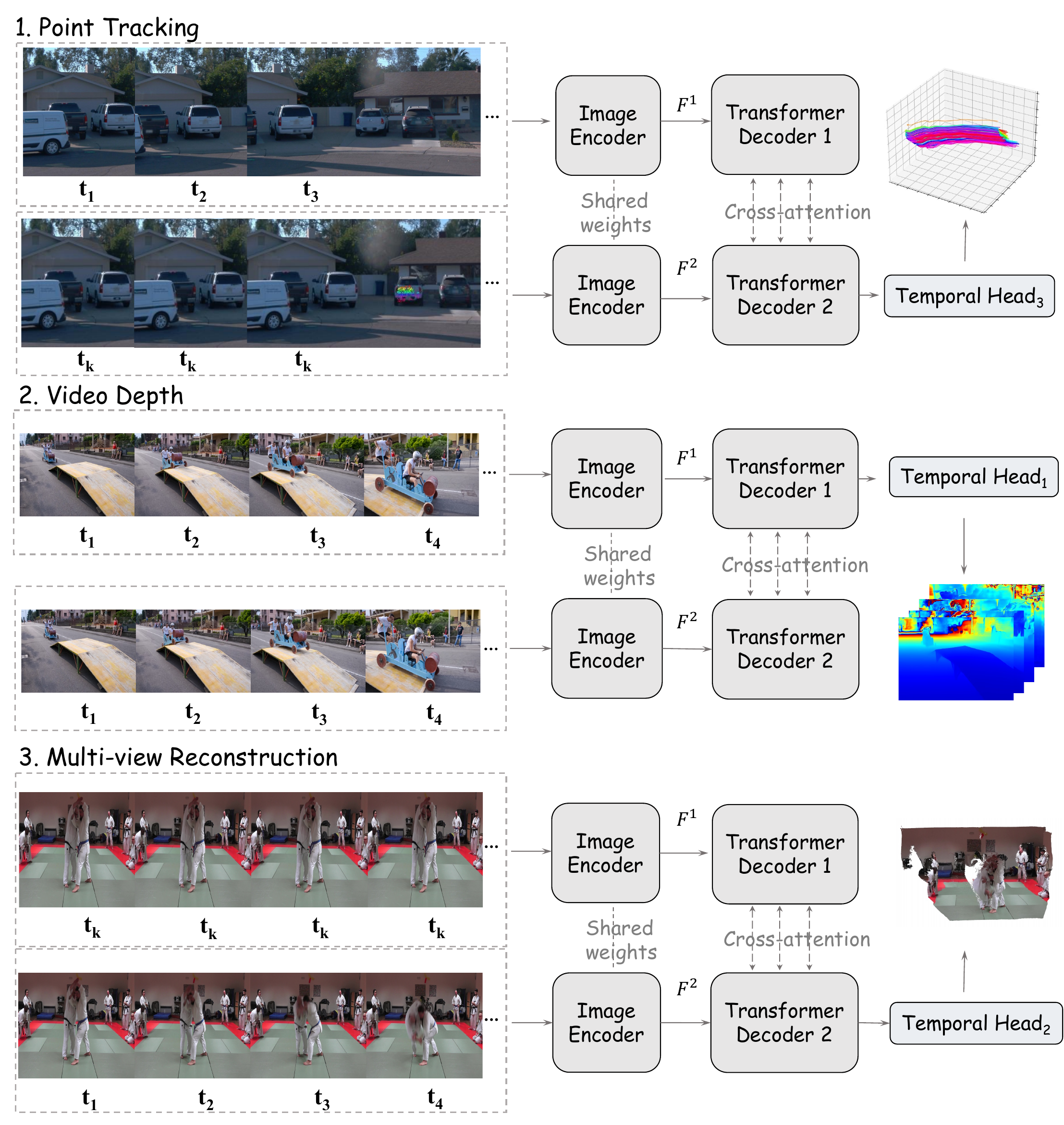}
    \caption{\textbf{Inference pipelines for point tracking, video depth, and multi-view reconstruction.} $t_k$ indicates the keyframe. With the help of the motion module and flexible input construction, POMATO can be easily applied to downstream temporal tasks.}
    \vspace{-4mm}
    \label{fig:inference_pipeline}
\end{figure}

Given a video sequence of $T$ frames ${\bI^{t_1}, \bI^{t_2}, \ldots, \bI^{t_T}}$, we construct a unique set of stereo image pairs for each task. As illustrated in Fig.~\ref{fig:inference_pipeline}, the flexible construction of input pairs—combined with the proposed temporal motion module and pointmap matching head—enables \OURS to seamlessly address downstream temporal tasks, including 3D point tracking, video depth estimation, and 3D reconstruction. The keyframe selection strategy and input formulation for each task are detailed in the following section. 

Besides the default regression losses for Head$_1$ and Head$_2$, and predefined losses Eq.~\ref{eq:regression} and Eq.~\ref{eq:matching_confidence} for Head$_3$, we further employ a temporal consistency loss, $\mathcal{L}_\text{t}$, which will be described in detail below.

In addition to the default regression losses for Head$_1$ and Head$_2$, and the predefined losses in Eq.~\ref{eq:regression} and Eq.~\ref{eq:matching_confidence} for Head$_3$, we further introduce a temporal consistency loss, $\mathcal{L}_\text{t}$, which will also be described in detail below.

\noindent\textbf{3D Point Tracking.}
As illustrated at the top of Fig.\ref{fig:inference_pipeline}, the keyframe is set to the first image of the global video sequence and fed to the proposed Head$_3$ to obtain the pointmap matching result of each query point (initialized at the first image) under the coordinate system of each reference frame $\{\bX^{t_1, t_1}_m,\bX^{t_1, t_2}_m,\bX^{t_1, t_3}_m,...\bX^{t_1, t_T}_m\}$, while the set of reference frames $\{\bI^{t_1},\bI^{t_2},\bI^{t_3},...\bI^{t_T}\}$ is fed to the Head$_1$ to obtain the pointmap under each ego coordinate system. The dense tracking results can be further sparsified by indexing the 2D coordinates. When inference on a video longer than $T$ frames, a simple sliding-window approach with an overlap of four frames is adopted to enhance the consistency between adjacent video windows. The temporal consistency loss $\mathcal{L}_\text{t}$ for tracking is:
{\small
\begin{equation}
\mathcal{L}_\text{t} = \frac{1}{T} \sum_{i=1}^T 
\left\Vert \frac{{\bX^{t_1,t_i}_m}}{z_{m}^{T}} 
- \frac{{\bar{\bX}^{t_1,t_i}}_{m}}{\bar{z}_{m}^{T}} \right\Vert 
+ \left\Vert \frac{{\bX^{t_i,t_i}}}{z^{T}} 
- \frac{{\bar{\bX}^{t_i,t_i}}}{\bar{z}^{T}} \right\Vert,
\label{eq:tracking_loss}
\end{equation}
}

where $z_{m}^T=$ norm $(\bX^{t_1,t_1}_m,\bX^{t_1,t_2}_m, ..., \bX^{t_1,t_T}_m)$ and $\bar{z}_{T}=$ norm $(\bar{\bX}^{t_1,t_1}_m,\bar{\bX}^{t_1,t_2}_m,..., \bar{\bX}^{t_1,t_T}_m)$. $z_{m}^T$ and $\bar{z}_{T}$ are defined analogously. 

\noindent\textbf{Video Depth Estimation.} As shown in the middle part of the Fig. \ref{fig:inference_pipeline}, the input video sequence is formulated to a set of identical image pairs $\{(\bI^{t_1}, \bI^{t_1}), (\bI^{t_2}, \bI^{t_2}),..., (\bI^{t_T}, \bI^{t_T})\}$ and fed to Head$_1$ and Head$_2$, where the predictions from each head are identical:$\{\bX^{t_1,t_1},\bX^{t_2,t_2},...,\bX^{t_N,t_N}\}$. We use the output of Head$_1$ as our final video depth estimation. The temporal consistency loss $\mathcal{L}_\text{t}$ is defined as:
{\small
\begin{equation}
\mathcal{L}_\text{t} = \frac{1}{T} \sum_{i=1}^T 
\left\Vert \frac{{\bX^{t_i,t_i}_1}}{z_{1}^{T}} 
- \frac{{\bar{\bX}^{t_i,t_i}}}{\bar{z}^{T}} \right\Vert 
+ \left\Vert \frac{{\bX_2^{t_i,t_i}}}{z_2^{T}} 
- \frac{{\bar{\bX}^{t_i,t_i}}}{\bar{z}^{T}} \right\Vert,
\label{eq:tracking_loss}
\end{equation}
}
where ${\bX^{t_i,t_i}_1}$ and ${\bX^{t_i,t_i}_2}$ indicate the output from Head$_1$ and Head$_2$, respectively. $\bar{\bX}^{t_i,t_i}$ is the pointmap groundtruth.

\noindent\textbf{3D Reconstruction}. Assisted by the temporal motion module, redundant post-process operations such as global alignment can be omitted, allowing the reconstructed 3D point cloud to be obtained in a feed-forward manner. As shown in the bottom part of Fig.\ref{fig:inference_pipeline}, the keyframe is set to the last frame $\bI^{t_T}$ within the temporal window of length $T$ and is fed to Head$_1$ with a set output of $\{\bX^{t_T, t_T},\bX^{t_T, t_T},...,\bX^{t_T, t_T}\}$. All the reference frames are input to the Head$_2$ so the target pointmaps $\{\bX^{t_1, t_T},\bX^{t_2, t_T},...,\bX^{t_T, t_T}\}$ are aligned under the coordinate system of the keyframe. The temporal consistency loss $\mathcal{L}_\text{t}$ is:
{\small
\begin{equation}
\mathcal{L}_\text{t} = \frac{1}{T} \sum_{i=1}^T 
\left\Vert \frac{{\bX^{t_T,t_T}}}{z^{T}_1} 
- \frac{{\bar{\bX}^{t_T,t_T}}}{\bar{z}_1^{T}} \right\Vert 
+ \left\Vert \frac{{\bX^{t_i,t_T}}}{z_2^{T}} 
- \frac{{\bar{\bX}^{t_i,t_T}}}{\bar{z}_2^{T}} \right\Vert
\label{eq:tracking_loss}
\end{equation}
}
We further freeze the parameters in Decoder1 and Decoder2 when training the temporal downstream tasks at the second stage. In our work, the temporal window length $T$ is set to 12. Additional explorations on the temporal length can be found in Sec.\ref{sec:exp}.

\begin{table*}[t]
\centering
\renewcommand{\arraystretch}{1.02}
\renewcommand{\tabcolsep}{1.5pt}
\resizebox{0.9\textwidth}{!}{
\begin{tabular}{@{}llcc>{\centering\arraybackslash}p{1.5cm}>{\centering\arraybackslash}p{1.5cm}|>{\centering\arraybackslash}p{1.5cm}>{\centering\arraybackslash}p{1.5cm}|>{\centering\arraybackslash}p{1.5cm}>{\centering\arraybackslash}p{1.5cm}@{}}
\toprule
 &  &  &  & \multicolumn{2}{c}{\textbf{Sintel}~\cite{sintel}} & \multicolumn{2}{c}{\textbf{BONN}~\cite{bonn}} & \multicolumn{2}{c}{\textbf{KITTI}~\cite{kitti}} \\ 
\cmidrule(lr){5-6} \cmidrule(lr){7-8} \cmidrule(lr){9-10}
\textbf{Alignment} & \textbf{Method} & \textbf{Optim.} & \textbf{Onl.\ } & {Abs Rel $\downarrow$} & {$\delta$\textless $1.25\uparrow$} & {Abs Rel $\downarrow$} & {$\delta$\textless $1.25\uparrow$} & {Abs Rel $\downarrow$} & {$\delta$ \textless $1.25\uparrow$} \\ 
\midrule

\multirow{6}{*}{\begin{minipage}{2cm}Per-sequence scale\end{minipage}} & DUSt3R-GA~\cite{dust3r} &  \checkmark & & 0.656 & {45.2} & {0.155} & {83.3} & {0.144} & {81.3}  \\
& MASt3R-GA~\cite{mast3r} & \checkmark & & 0.641 & {43.9} & {0.252} & {70.1} & {0.183} & {74.5} \\

& MonST3R-GA~\cite{monst3r} &  \checkmark &  & \textbf{0.378} & \textbf{55.8} & \textbf{0.067} & \textbf{96.3} & {0.168} & {74.4}  \\
& Spann3R~\cite{spann3r} &  & \checkmark & 0.622 & {42.6} & {0.144} & {81.3} & {0.198} & {73.7}  \\

& CUT3R~\cite{cut3r} &  & \checkmark & {0.421}  & {47.9} & {0.078} & {93.7} & \underline{0.118} & \underline{88.1}  \\
& \OURS & & \checkmark  & \underline{0.416} & \underline{53.6} & \underline{0.074} & \underline{96.1} & \textbf{0.085} & \textbf{93.3} \\
\midrule

 \multirow{3}{*}{\begin{minipage}{2cm}Per-sequence scale \& shift\end{minipage}} &  MonST3R-GA~\cite{monst3r} & \checkmark  & & \textbf{0.335} & \textbf{58.5} & \textbf{0.063} & 96.4 & {0.104} & {89.5}   \\
 & CUT3R~\cite{cut3r} &  & \checkmark & 0.466 & 56.2 & 0.111 & 88.3 & \textbf{0.075} & \textbf{94.3}   \\
& \OURS & & \checkmark & \underline{0.345} & \underline{57.9} & \underline{0.072} & \textbf{96.5} & \underline{0.084} & \underline{93.4} \\
\bottomrule
\end{tabular}
}
\vspace{-.05in}
\caption{\small{
\textbf{Video depth evaluation}. We report scale-invariant depth and scale $\&$ shift invariant depth accuracy on Sintel~\cite{sintel}, Bonn~\cite{bonn}, and KITTI~\cite{kitti} datasets. Methods requiring global alignment are marked ``GA'', while ``Optim.'' and ``Onl.'' indicate optimization-based and online methods, respectively. The best and second best results in each category are \textbf{bold} and \underline{underlined}, respectively.}
}
\label{tab:video_depth}
\end{table*}

\section{Experiments}
\label{sec:exp}

\subsection{Experimental Details}
\noindent\textbf{Training data.} We train our network with a mixture of five datasets: PointOdyssey~\cite{pointodyssey}, Tartanair~\cite{tartanair}, ParallelDomain4D~\cite{van2024generative}, DynamicReplica~\cite{karaev2023dynamicstereo} and Carla (0.9.15)~\cite{carla}. The specific number and the usage ratio of each dataset can be found in the supplementary materials. All datasets include pixel-accurate ground truth depth, as well as camera intrinsics and extrinsics, and encompass a wide variety of dynamic scenes across both indoor and outdoor environments. Among them, PointOdyssey and DynamicReplica have additional 2D trajectory annotations for dynamic objects which can be used to construct pointmap matching ground truth following Eq. \ref{eq:def_pointmap_matching}. All datasets are used to supervise geometry learning on Head$_1$ and Head$_2$, while only PointOdyssey, DynamicReplica, and TartanAir are used to train the proposed pointmap matching head.

\noindent\textbf{Training and inference details.} Our model architecture is based on the publicly available DUSt3R~\cite{monst3r} model, utilizing the same backbone consisting of a ViT-Large encoder and a ViT-Base decoder. To fully leverage \monster's geometry estimation capabilities in dynamic scenes, we initialize our model using the publicly available \monster{} checkpoint. For the newly introduced pointmap matching head, we initialize its weights from the pretrained Head$_2$ weights of \monster. The temporal motion module is initialized following \cite{animatediff}. We train our network for 10 epochs with a cosine learning rate schedule, with an initial learning rate of 1e-4. In the first stage, which involves pairwise training, we use a batch size of 16 on 4 A100 GPUs (40G). In the second stage, where the temporal motion module is introduced, the batch size is set to 4 with a fixed temporal window length of 12. During each training iteration, we randomly sample a downstream task—3D point tracking, video depth estimation, or 3D reconstruction—to construct the input pairs and apply the corresponding loss function.

\subsection{Video Depth Estimation}

Following MonST3R~\cite{monst3r} and CUT3R~\cite{cut3r}, we rescale all predictions from the same video to align them together by conducting two forms of alignment: per-sequence scale and shift alignment and per-sequence scale alignment. Thus, we can measure the per-frame depth quality and inter-frame depth consistency. We employ our proposed motion module for video depth estimation in a feed-forward manner as described in Sec.\ref{sec:temporal_tasks} and compare our method against several variants of DUSt3R, including DUSt3R~\cite{dust3r}, MAST3R~\cite{mast3r}, MonST3R~\cite{monst3r}, Spann3R~\cite{spann3r}, and CUT3R~\cite{cut3r}. Given 6 frames of 288×512 on an NVIDIA 4070 GPU, \OURS reconstructs the 3D point cloud in 0.7 seconds, whereas global alignment-based methods such as MonST3R require 5.8 seconds. As shown in Tab. \ref{tab:video_depth}, our method demonstrates comparable performance to the global alignment (GA)-based MonST3R~\cite{monst3r} on the Sintel~\cite{sintel} and BONN~\cite{bonn} datasets, while surpassing it on KITTI dataset. Besides, we consistently outperform the state-of-the-art online method, CUT3R~\cite{cut3r}, across various settings. These results underscore the effectiveness of our approach, specifically (1) the joint learning of geometry and pointmap matching, and (2) the temporal motion module.

\begin{table*}[!t]
\centering
\footnotesize
\renewcommand{\arraystretch}{1.0}
\renewcommand{\tabcolsep}{8pt}

\aftertabcaption
\resizebox{0.72\textwidth}{!}{
\begin{tabular}{rcc|cc|cc|cc}
\toprule
  & \multicolumn{2}{c}{PointOdyssey~\cite{pointodyssey}} & \multicolumn{2}{c}{ADT~\cite{aria}} & \multicolumn{2}{c}{PStudio~\cite{pstudio}} & \multicolumn{2}{c}{\textbf{Average}} \\ 
 
\cmidrule(lr){2-3} \cmidrule(lr){4-5} \cmidrule(lr){6-7} \cmidrule(lr){8-9}
 {Method} & {L-12} & {L-24} & {L-12} & {L-24} & {L-12} & {L-24} & {L-12} & {L-24} \\ 
\midrule
SpatialTracker$^*$~\cite{spatialtracker} & 20.46 & 20.71 & 21.64 & 20.67 & \textbf{30.41} & \textbf{25.87} & \underline{24.17} & \underline{22.42 }\\  \midrule

DUSt3R~\cite{dust3r} & 19.03 & 19.03 & 29.02 & 25.55 & 9.72 & 6.50 & 19.26 & 17.03 \\ 
MASt3R~\cite{mast3r} & 16.58 & 17.35 & 27.36 & 26.46 & 11.78 & 8.09 & 18.57 & 17.30 \\
MonST3R~\cite{monst3r} & \underline{27.31} & \underline{27.92} & \underline{28.30} & \underline{26.13} & 16.50 & 11.06 & 24.03& 21.70 \\ 
\OURS & \textbf{33.20} & \textbf{33.58} & \textbf{31.57} & \textbf{28.22} & \underline{24.59} & \underline{19.79} & \textbf{29.79}& \textbf{27.20} \\
\bottomrule
\end{tabular}
}
\aftertab
\caption{\textbf{3D tracking evaluation.} We report the APD metric to evaluate 3D point tracking on the PointOdyssey~\cite{pointodyssey}, ADT~\cite{aria}, and PStudio~\cite{pstudio} datasets. L-12 and L-24 indicate tracking within the temporal length of 12 frames and 24 frames, respectively.}
\label{tab:tracking}
\end{table*}

\begin{figure*}[!t]
    \centering
\includegraphics[width=0.95\linewidth]{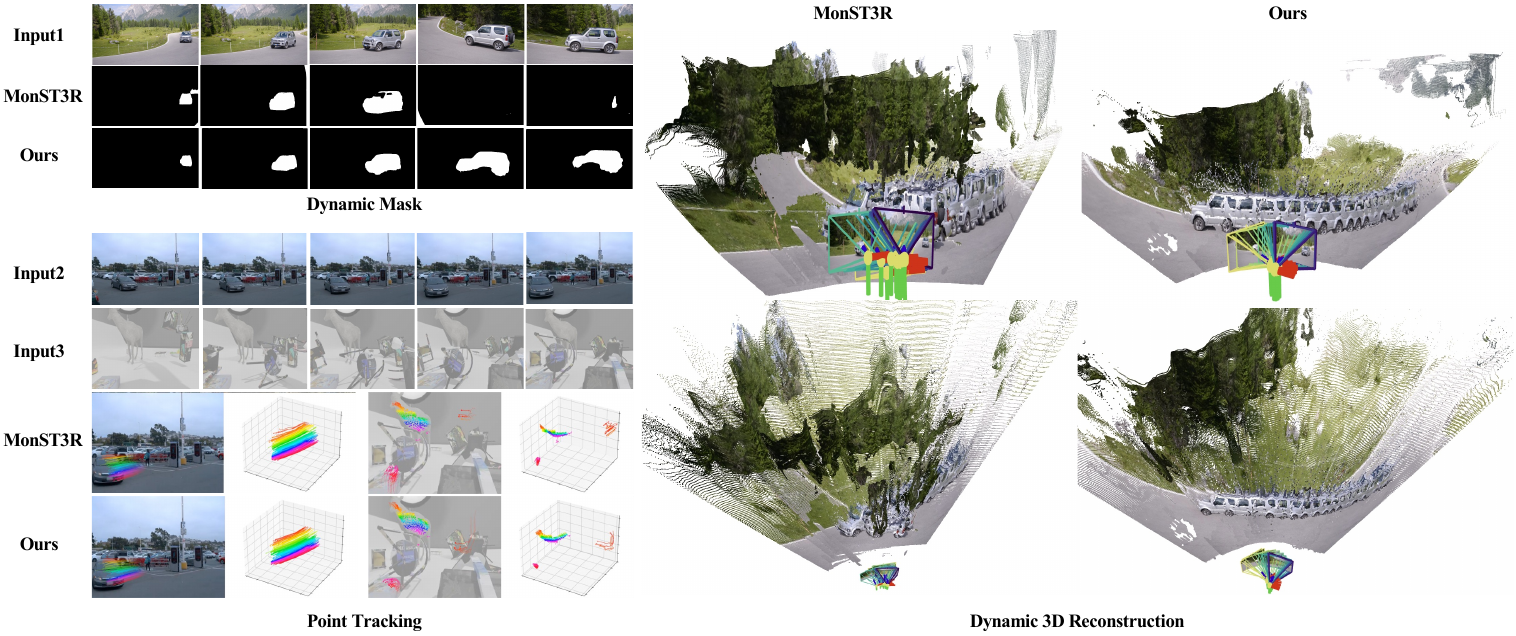}
    \caption{\textbf{Qualitative comparison of dynamic scenes.} Compared to \monster{}, our \OURS can provide more reliable motion masks, 3D point tracking, and reconstruction performance.}
    \label{fig:vis_motion}
\end{figure*}

\subsection{3D Point Tracking}
For 3D point tracking task, we use the Aria Digital Twin (ADT)~\cite{aria}, and Panoptic Studio (PStudio)~\cite{pstudio} benchmarks from the TAPVid-3D~\cite{TAPVid3D} dataset along with the validation set on the PointOdyssey~\cite{pointodyssey} dataset. We report the Average Percent Deviation (APD) metric, which quantifies the average percentage of points within a threshold relative to the ground truth depth. The APD metric serves as a direct measure of the accuracy of the predicted tracking. We reformulate the datasets and project all the query points within a temporal window to the first frame. We report tracking results on the length of 12 and 24 frames. As shown in Tab.\ref{tab:tracking}, our \OURS achieves the best performance on both PointOdyssey and ADT datasets. It's worth mentioning that SpatialTracker \cite{spatialtracker} is a state-of-the-art network tailored for 3D point tracking with ground truth camera intrinsic as additional input data. \OURS surpasses it on two datasets and improves the average APD metric by 23.3\% and 21.4\% for 12 frames and 24 frames, respectively. For DUSt3R-based methods, we use the output of Head$_2$ as tracking results. Obviously, the ambiguous matching representation limits its capability to handle this fine-grained 3D reconstruction task in dynamic scenes.

\begin{table*}[!ht]
\centering
\footnotesize
\renewcommand{\arraystretch}{1.0}
\renewcommand{\tabcolsep}{1pt}
\label{tab:temporal_ablation}
\resizebox{0.8\textwidth}{!}{
\begin{tabular}{@{}c|cc|cc|cc|ccc@{}}
\toprule
 & \multicolumn{6}{c|}{Video Depth} & \multicolumn{3}{c}{Tracking (12 Frames)} \\ 
\cmidrule(lr){2-7} \cmidrule(lr){8-10}

{Temporal Length} & \multicolumn{2}{c|}{Sintel~\cite{sintel}} & \multicolumn{2}{c|}{Bonn~\cite{bonn}} & \multicolumn{2}{c|}{KITTI~\cite{kitti}} & PointOdyssey~\cite{pointodyssey} & ADT~\cite{aria} & PStudio~\cite{pstudio}  \\
\cmidrule(lr){2-7} \cmidrule(lr){8-10}

 & {Abs Rel $\downarrow$} & {$\delta$\textless $1.25\uparrow$} & {Abs Rel $\downarrow$} & {$\delta$\textless $1.25\uparrow$} & {Abs Rel $\downarrow$} & {$\delta$\textless $1.25\uparrow$} & APD$\uparrow$ & APD$\uparrow$ & APD$\uparrow$  \\ \midrule
Pair-wise & 0.548 & 46.2 & 0.087 & 94.0 & 0.113 & 89.5 & 32.06 & 29.87 & 23.10\\
6 frames & \underline{0.436} & \underline{51.3} & \underline{0.076} & \underline{95.9} & \textbf{0.085} & \textbf{93.5} & \underline{32.69} & \underline{30.93}& \underline{24.52}\\
12 frames  & \textbf{0.416} & \textbf{53.6} & \textbf{0.075 }& \textbf{96.1} & \underline{0.086} & \underline{93.3} & \textbf{33.20} & \textbf{31.57}& \textbf{24.59}\\
\bottomrule
\end{tabular}
}
\caption{\textbf{Ablation study on the temporal motion module}. The introduction of the temporal motion module brings a significant improvement. As the temporal window length enlarges from 6 frames to 12 frames, we obtain an overall consistent improvement.}
\label{tab_ablation}
\aftertabcaption
\aftertab
\end{table*}
\begin{table}[!t]
\centering
\footnotesize
\renewcommand{\arraystretch}{0.95}
\renewcommand{\tabcolsep}{2.pt}
\label{tab:temporal_length}
\resizebox{0.45\textwidth}{!}{
\begin{tabular}{lccc|ccc}
\toprule
  & \multicolumn{3}{c}{TUM~\cite{tum}} & \multicolumn{3}{c}{Bonn~\cite{bonn}} \\
\cmidrule(lr){2-4} \cmidrule(lr){5-7} 
 Method & {ATE $\downarrow$} & {RPE trans $\downarrow$} & {RPE rot $\downarrow$} & {ATE $\downarrow$} & {RPE trans $\downarrow$} & {RPE rot $\downarrow$} \\
\midrule
DUSt3R~\cite{dust3r} & 0.025 & 0.013 & 2.361 & 0.030 & 0.025 & 2.522 \\
MASt3R~\cite{mast3r} & 0.027 & 0.015 & 1.910 & 0.031 & 0.025 & 2.478 \\

MonST3R~\cite{monst3r} & \underline{0.021} & \textbf{0.006} & 1.142 & \textbf{0.025} & \underline{0.021} & \underline{2.120}\\
CUT3R~\cite{cut3r} & 0.023 & 0.016 & \underline{0.510} & \underline{0.028} & 0.033 & 2.569 \\

\OURS & \textbf{0.020} & \underline{0.010} & \textbf{0.509} & 0.037 & \textbf{0.016} &\textbf{1.782}  \\ \bottomrule

\end{tabular}
}
\caption{\textbf{Pose estimation}. Our method achieves an overall best performance and improves the RPE rot metric significantly.}
\label{tab:camera_pose}
\aftertabcaption
\aftertab
\end{table}

\begin{figure}[!t]
    \centering
\includegraphics[width=0.95\linewidth]{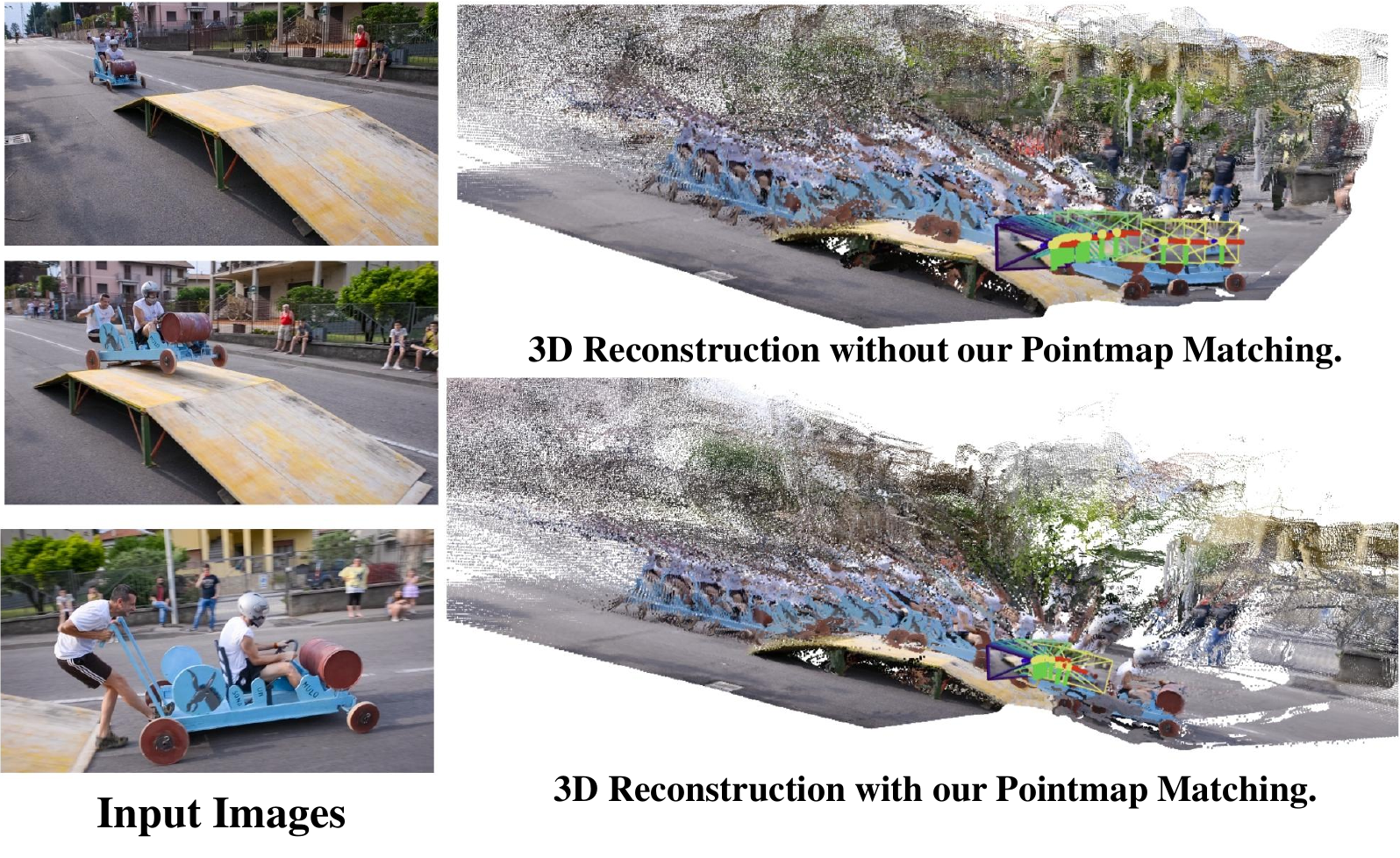}
    \caption{\textbf{Effectiveness of our pointmap matching head}. Without explicitly filtering out the motion area, both pose and geometry estimation will be degraded.}
    \label{fig:head3_pose}
\end{figure}

\subsection{Camera Pose Estimation}
Following DUSt3R-based methods, we perform global alignment with the model trained in the first stage on the Bonn~\cite{bonn} and TUM~\cite{tum} datasets. The sampling stride is set to 5 for the Bonn dataset and 3 for the TUM dataset. Compared with optical-flow assisted global alignment in MonST3R, the dynamic mask is computed according to Eq.~\ref{eq:dynamic_mask_eq} while the 2D pseudo label is replaced by projecting the pointmap matching results to 2D coordinates with estimated camera intrinsic. Absolute Translation Error (ATE), Relative Translation Error (RPE trans), and Relative Rotation Error (RPE rot) are reported. The evaluation results over 40 frames are reported in Tab.~\ref{tab:camera_pose}. Notably, \OURS obtains an overall state-of-the-art performance and significantly improves the RPE-rot metric, surpassing MonST3R by 55.4\% and 13.3\% on the TUM and Bonn datasets. 

\subsection{Ablation Study}
\begin{table}[!t]
\centering
\footnotesize
\renewcommand{\arraystretch}{1.0}
\renewcommand{\tabcolsep}{2.pt}
\label{tab:temporal_length}
\resizebox{0.45\textwidth}{!}{
\begin{tabular}{lccc|ccc}
\toprule
  & \multicolumn{3}{c}{Bonn~\cite{bonn}} & PointOdyssey~\cite{pointodyssey} & ADT~\cite{aria} & PStudio~\cite{pstudio} \\
\cmidrule(lr){2-4} \cmidrule(lr){5-7} 
 Method & {ATE $\downarrow$} & {RPE trans $\downarrow$} & {RPE rot $\downarrow$} & {APD $\uparrow$} & {APD $\uparrow$} & {APD $\uparrow$} \\
\midrule

 W/O Head$_3$ & 0.040 & \textbf{0.015} & \textbf{1.721} & 29.10 & 29.62 &16.94 \\ 
 W/ Head$_3$ & \textbf{0.037} & 0.016 & 1.782 & \textbf{32.06} & \textbf{29.87} & \textbf{23.10} \\ \bottomrule
 
\end{tabular}
}
\caption{\textbf{Ablation study on the effectiveness of the pointmap matching head}. The comparisons are reported on the pose estimation and 3D point tracking tasks.}
\label{tab:head3_ablation}
\aftertabcaption
\aftertab
\end{table}

We conduct extensive ablation studies to evaluate the effectiveness of the temporal motion module and the proposed pointmap matching head. As shown in Table~\ref{tab_ablation}, we report results for three models: one trained with only pairwise images (first-stage training), one using a shorter temporal window of 6 frames, and another using the default temporal window length of 12 frames. Incorporating temporal consistency yields substantial improvements across all datasets for video depth estimation and 3D point tracking. Further improvement is achieved when the temporal window length increases from 6 frames to 12 frames. In Table~\ref{tab:head3_ablation}, we evaluate the effectiveness of the pointmap matching head. While it introduces only a modest improvement in the ATE metric, we attribute this to the limited motion and minimal viewpoint variation in the indoor evaluation dataset. As illustrated in Fig.~\ref{fig:head3_pose}, under challenging in-the-wild conditions with significant motion and rapid viewpoint changes, removing the pointmap matching head introduces ambiguity in explicit rigid transformation estimation, resulting in a clear degradation in performance. To further demonstrate the impact of the pointmap matching head on 3D point tracking, we conduct tracking experiments over 12 frames using the pairwise input setup. Clearly, removing the pointmap matching head (using only Head$_2$) leads to an inevitable performance drop, emphasizing explicit correspondence modeling for reliable long-term tracking.

\section{Discussion and Conclusion}

We introduce \OURS, a unified framework for geometry estimation and motion understanding in dynamic scenes. By leveraging the proposed pointmap matching head, our method effectively distinguishes moving regions, thereby mitigating the interference introduced by dynamic objects. The temporal motion module further facilitates the learning of temporal dynamics across frames, enhancing scale consistency and improving performance in tasks where both geometry and matching are critical—most notably, 3D point tracking. The downstream temporal tasks including 3D point tracking, video depth estimation, and 3D reconstruction can be easily applied in a feed-forward manner. In future work, we plan to scale up training with more dynamic reconstruction and matching datasets to further enhance 3D reconstruction and tracking performance.

\textbf{Acknowledgement}. This work was supported by the National Natural Science Foundation of China (No. 62206244)
\newpage
{
    \small
    \bibliographystyle{ieeenat_fullname}
    \bibliography{main}
}

\clearpage
\appendix
\setcounter{page}{1}
\maketitlesupplementary
\begin{table*}[!t]
\centering
\footnotesize
\renewcommand{\arraystretch}{0.95}
\renewcommand{\tabcolsep}{2.pt}
\resizebox{0.85\textwidth}{!}{
\begin{tabular}{l|c|c|c|c|c|c}
\toprule
Dataset & Domain & Scene Type & \# of Frames & \# of Scenes & Dynamics & Ratio \\ 
\midrule
PointOdyssey~\cite{pointodyssey} & Synthetic & Indoors \& Outdoors & 200k & 131 & Realistic & 57.1\% \\
TartanAir~\cite{tartanair} & Synthetic & Indoors \& Outdoors & 100k & 163 & None & 14.3\% \\
DynamicReplica~\cite{karaev2023dynamicstereo} & Synthetic & Indoors & 145k & 524 & Realistic & 14.3\%\\ 
ParallelDomain4D~\cite{van2024generative} & Synthetic & Outdoors & 750k & 15015 & Driving & 8.6\% \\
Carla~\cite{carla} & Synthetic & Outdoors & 7k & 5 & Driving & 5.7\%\\
\bottomrule
\end{tabular}\textbf{}
}
\caption{\textbf{An overview of all training datasets and sample ratio.} All datasets provide both camera pose, depth, and most of them include dynamic objects.}
\label{tab:train_data}
\aftertabcaption
\aftertab
\end{table*}

\begin{figure*}[ht]
    \centering
    \includegraphics[width=.99\linewidth]{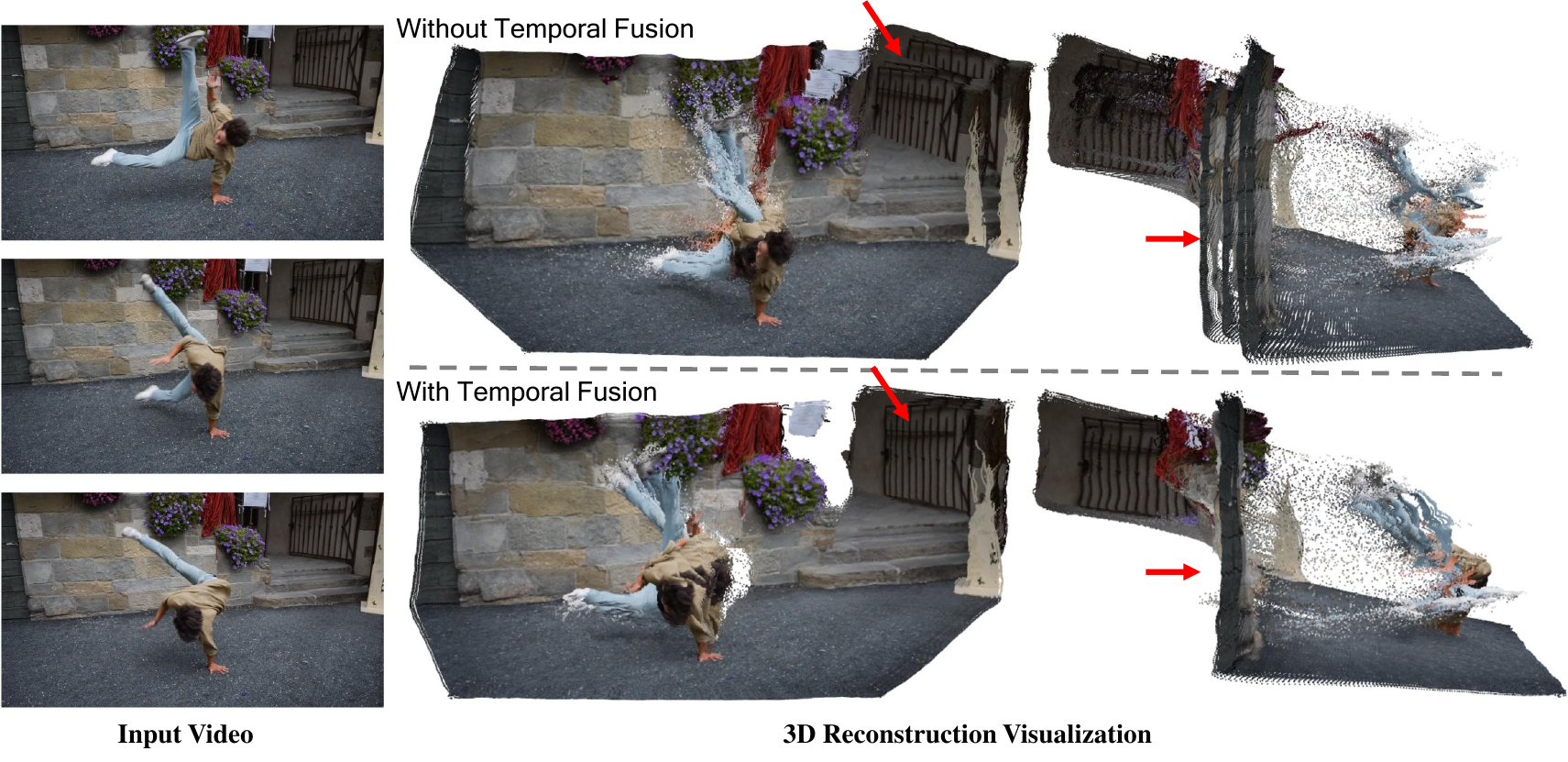}
    \caption{\textbf{Fast 3D reconstruction with our temporal motion module}. Given a sequence of images less than temporal window length, our \OURS can directly obtain a global pointmap under the key frame coordinate.}
    \vspace{-4mm}
    \label{fig:temporal_recon}
\end{figure*}

\section{Pointmap Matching for Global Alignment.}
Given a sequence of video frames, the target of global alignment is to project all pairwise estimated pointmaps to the same global world coordinates. DUSt3R constructs a connectivity pairwise graph and aims to minimize the re-projection error for each image pair globally where the dynamic regions are supposed to be separated from the static regions. To this end, MonST3R \cite{monst3r} further introduces an assistant optical flow network \cite{raft2} to help mask the dynamic regions and provide a pseudo label of 2D matching for minimizing the re-projection error in static regions. However, the introduced assistant model will introduce inevitable domain gaps and additional computation costs. Besides, the optical flow model is tailored for matching within two adjacent frames, suffering an obvious degeneration with the large view displacement. In \OURS, for an image pair $\{\bI^i, \bI^j\}$, the dynamic mask $\bD^{j,i}$ is calculated by comparing the difference between $\bX^{j,i}$ and $\bX^{j,i}_m$:
\begin{align}
    \bD^{j,i} = ||\bX^{j,i}_m - \bX^{j,i}|| > \alpha,
\end{align}
where $\alpha$ is a dynamic threshold defined as $3 \times \text{median}(||\bX^{j,i}_m - \bX^{j,i}||)$.

Given the updated camera intrinsic $\tilde K$ after an iteration of optimization, the target matching 2D coordinates $\bF^{j,i}_m \in \matR^{H\times W \times 2}$ can be calculated as $\bF^{j,i}_m = p(\tilde \bK \bX^{j,i}_m)$ where $p$ is a mapping from 3D camera coordinates to 2D pixel coordinates. The optical flow loss proposed in MonST3R can thus be modified with our dynamic mask and 2D matching coordinates. Details about the optical flow loss are referred to MonST3R \cite{monst3r}.

\section{Fast 3D Reconstruction with video \OURS}
Given a sequence of images less than the temporal window length of 12 frames, dynamic 3D reconstruction can be obtained by directly estimating the pointmaps of all reference images to the coordinate of the key frame as discussed in the Sec.\ref{sec:motion_module}. Here, we provide more visualization results of this feed-forward manner and demonstrate the effectiveness of introducing the temporal motion module. As shown in Fig.\ref{fig:temporal_recon}, directly applying the pairwise reconstruction will suffer from an obvious scale shift among different frames. After the temporal motion module, the consistency within the video sequence obtains an obvious enhancement. 

\section{Training Data Details}
The details about the training datasets can be found in Tab.\ref{tab:train_data}. The finetuning procedure of \OURS was conducted exclusively using synthetic training datasets.

\section{More Visualizations on Dynamic Scenes}
\begin{figure*}[hb]
    \centering
    \includegraphics[width=0.99\linewidth]{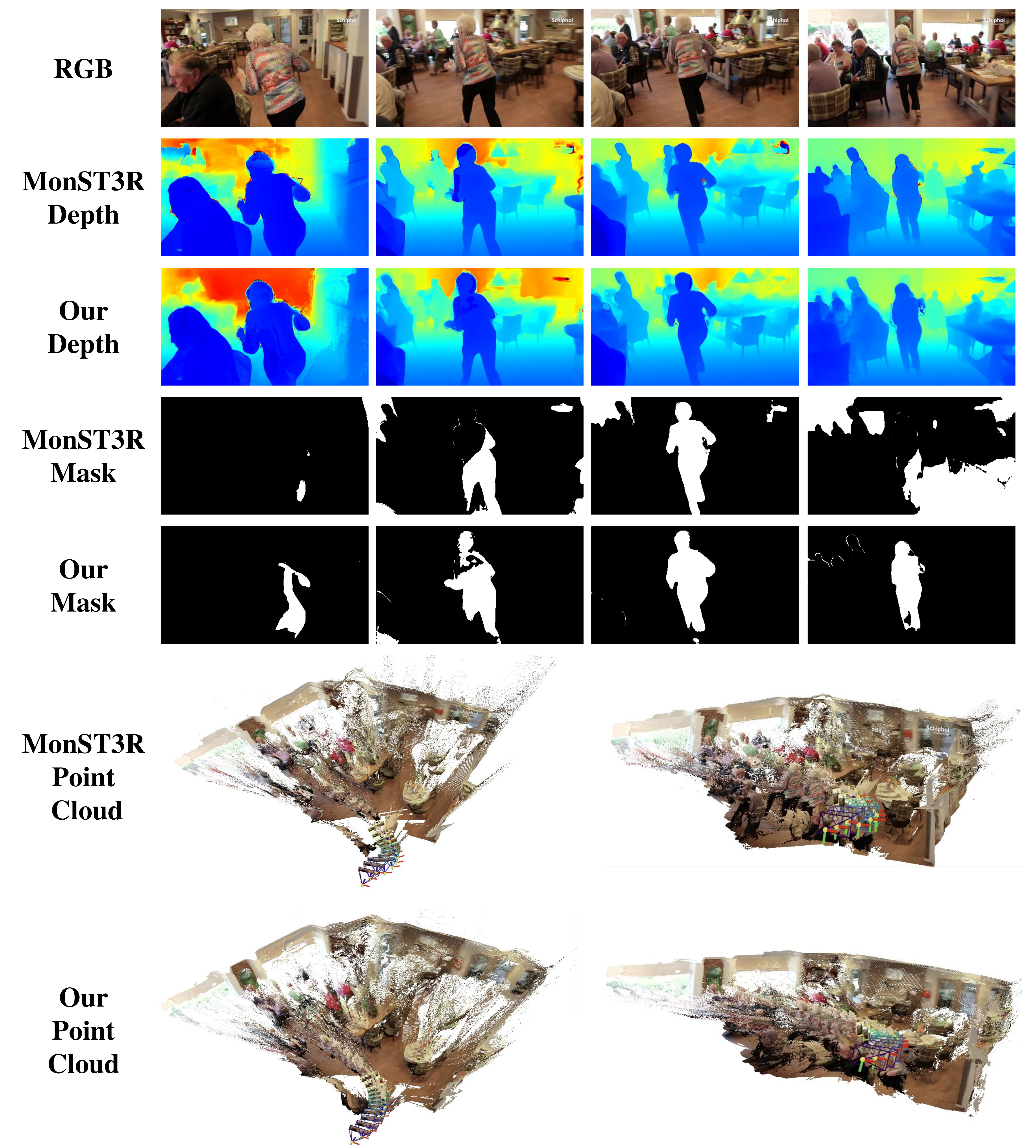}
    \caption{Compared with MonST3R, our \OURS can provide more complete dynamic masks and consistent geometry. }
    \label{fig:supp_vis1}
\end{figure*}

\begin{figure*}
    \centering
    \includegraphics[width=\linewidth]{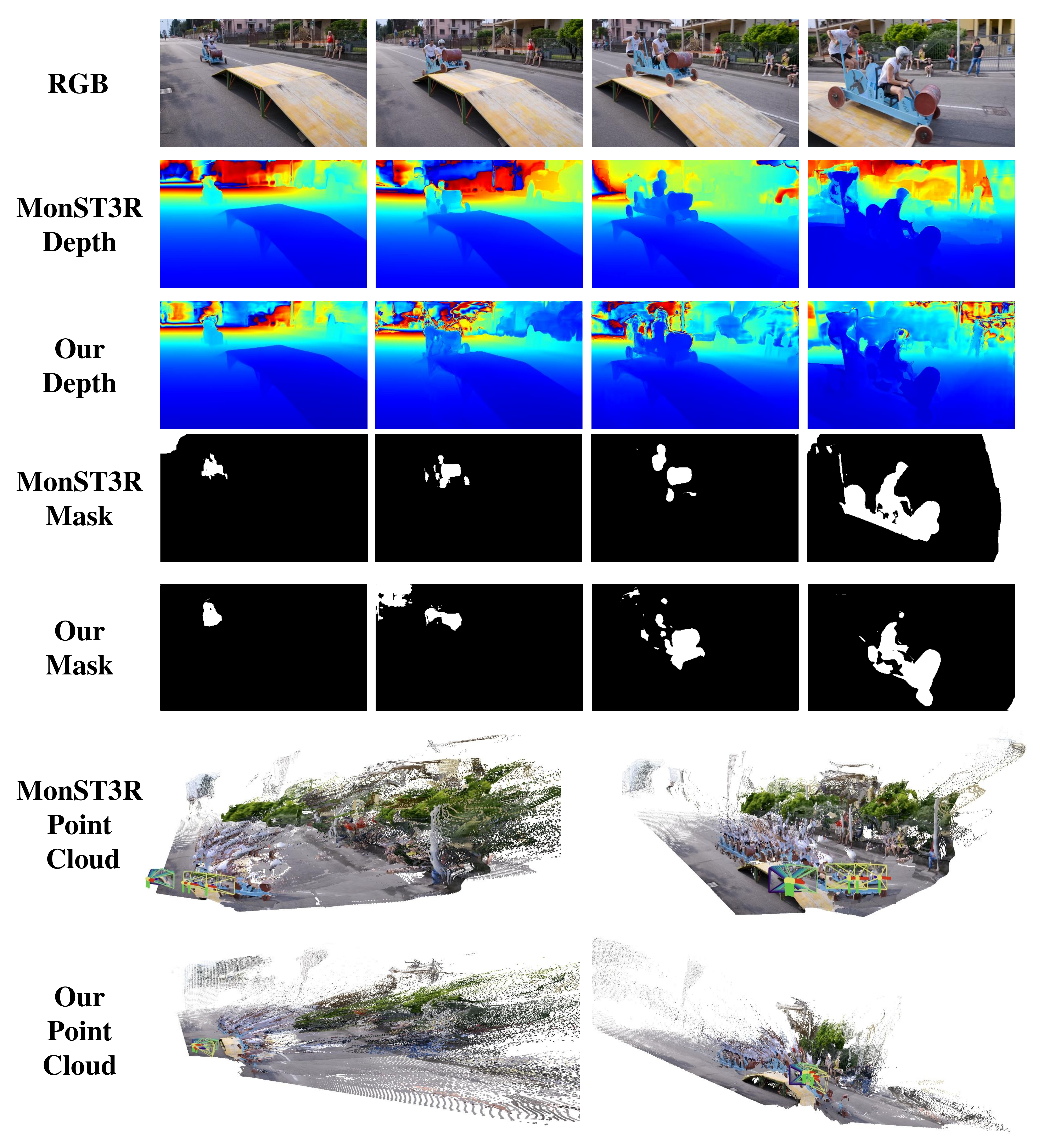}
    \caption{MonST3R suffers obvious degeneration when the view displacement is large as reflected by the erroneous pose estimation while \OURS can still provide a consistent camera trajectory.}
    \label{fig:supp_vis2}
\end{figure*}

We provide more visualizations in Fig. \ref{fig:supp_vis1} and Fig. \ref{fig:supp_vis2}. MonST3R suffers obvious degeneration when the view displacement is large as reflected by the erroneous pose estimation while
\OURS can still provide a consistent camera trajectory. 

\end{document}